 \documentclass[final,1p,times]{elsarticle}

\usepackage{amssymb}
\usepackage{epsfig}
\usepackage{longtable}
\usepackage{rotating}
\usepackage{lscape}
\usepackage{booktabs}
\usepackage[singlelinecheck=off]{caption}

\newcommand{\be}{\begin{equation}}
\newcommand{\ee}{\end{equation}}

\begin{document}
\begin{frontmatter}

\title{Direct local parametrization of nuclear state densities using the back-shifted Bethe formula}

\author[lanl]{C. \"{O}zen}
\author[yale]{Y. Alhassid\corref{cor1}}
\ead{yoram.alhassid@yale.edu}

\cortext[cor1]{Corresponding author}

\address[lanl]{Applied Physics (XTD) Division, MS K784, Los Alamos National Laboratory, Los Alamos, NM 87545, USA}
\address[yale]{Center for Theoretical Physics, Sloane Physics Laboratory, Yale University,\\ New Haven, CT 06520, USA}

\begin{abstract}
Level densities are often parametrized using the back-shifted Bethe formula (BBF) for nuclei that possess experimental data for s-wave neutron resonance average spacings and a complete discrete level sequence at low excitation energies. However, these parametrizations require the additional modeling of the dependence of  the spin-cutoff parameter on excitation energy. Here we avoid the need to model the spin distribution of level densities by using the experimental data to parametrize directly the state densities, for which the BBF does not depend on the spin-cutoff parameter.
This approach allows for a local parameterization of state densities that is independent of the spin-cutoff parameter.
We provide these parameters in a tabulated form for applications in nuclear reaction calculations and for testing microscopic approaches to state densities.
\end{abstract}

\end{frontmatter}

\section{Introduction}
\label{Sec:Intro}

Nuclear level densities (NLDs) are important input to the Hauser-Feshbach theory~\cite{Hauser1952}  of compound nucleus reaction rates~\cite{Burbidge1957,Rauscher2000, Koning2012}. 
They are often parametrized in terms of the back-shifted Bethe formula (BBF)~\cite{Huizenga1969, Huizenga1972}, also known as the back-shifted Fermi gas model, and assume a spin distribution which is described by the spin-cutoff model~\cite{Bethe1937,Ericson1960,Iljinov1992,Rauscher1997}.   However, this approach requires the modeling of the dependence of the spin-cutoff parameter, or alternatively the moment of inertia, on excitation energy.  Furthermore, pairing correlations in even-even nuclei cause deviations of the spin distribution from the spin-cutoff model at sufficiently low excitation energies~\cite{Alhassid2007,Alhassid2005}.

In this work, we parametrize directly the state densities for nuclei that possess neutron resonance level spacings data and have a complete sequence of low-lying energy levels together with their spins.  The advantage of this approach is that the BBF for the state density depends only on the single-particle level density parameter $a$ and the backshift parameter $\Delta$ but not on the spin-cutoff parameter $\sigma$.  Thus, modeling of the dependence of  $\sigma$ (or alternatively the nuclear moment of inertia) on excitation energy is not needed.

The outline of this paper is as follows:  In Sec.~2, we review Fermi gas models and the spin-cutoff model and discuss state densities vs.~level densities. In Sec.~3, we present the method we use to determine the BBF state density parameters from experimental data. In Sec.~4 we present our results for 294 nuclei for which neutron resonance data and complete set of low-lying levels exist. Sec.~5 presents our conclusions.

\section{Fermi gas models and spin-cutoff model}

\subsection{Fermi gas models}

The Fermi gas formula for the state density is based on a thermodynamic approach to non-interacting fermions in the grand-canonical ensemble, in which the state density is calculated from the grand-canonical partition function in the saddle-point approximation.  Assuming  two kinds of fermions (i.e., protons and neutrons), the Fermi gas state density is given by~\cite{Bethe1936}
\begin{eqnarray}
   \label{Eq:FG}
   \rho(E)=\frac{\sqrt{\pi}}{12}a^{-1/4}E^{-5/4}
   e^{2\sqrt{aE}} \;.
\end{eqnarray}

The parameter $a$ is given by $a=\frac{\pi^2}{6}[ g_p(\epsilon^{(p)}_F) + g_n(\epsilon^{(n)}_F)]$, where $g_p(\epsilon)$ is the single-particle level density of protons (neutrons) at energy $\epsilon$, and $\epsilon^{(p)}_F$ is the proton (neutron) Fermi energy.  This formula is derived assuming $Z \approx N$. 
For $Z \neq N$, the state density is given by an equation similar to Eq.~(\ref{Eq:FG}) but with an additional factor of $g/(2\sqrt{g_p\, g_n})$ on its r.h.s. This factor is of the order of unity.

It is a challenge to calculate the level density in the presence of correlations, and most theoretical estimates are based on empirical modifications of the Fermi gas model to account for correlations.   In particular, the BBF~\cite{Huizenga1969} is obtained from Bethe formula (\ref{Eq:FG}) by shifting the ground-state energy by $\Delta$
 \be\label{BBF}
 \rho(E_x) = \frac{\sqrt{\pi}}{12} a^{-1/4} (E_x-\Delta)^{-5/4} e^{2 \sqrt{a (E_x - \Delta)}}\;.
 \ee
 The parameter $\Delta$ is known as the backshift parameter.

Shell effects can be taken into account through an energy dependence of the corresponding model's parameters~\cite{Ignatyuk1975,Kataria1978}. The parameters are determined by fits of available experimental data in individual nuclei~\cite{Dilg1973,Iljinov1992,Koning2008}, which we will refer to as local fits, or by global fits~\cite{Rauscher1997,Koning2008}.

\subsection{Spin-cutoff model} 

The spin-cutoff model can be derived by the random coupling of the single-particle spins~\cite{Bethe1937,Ericson1960}. In this model, the distribution of the spin projection $M$ is a Gaussian
\be\label{M-dist}
\frac{\rho_M}{\rho} = \frac{1}{\sqrt{2\pi} \sigma}  e^{-M^2/2\sigma^2}\;,
\ee
where $\sigma=\langle J_z^2 \rangle^{1/2}$ is known as the spin-cutoff parameter.  Using the equipartition theorem at temperature $T$, we find
\be\label{sigma-I}
\sigma^2 = \frac{I T}{\hbar^2}\;,
\ee
where $T$ is the thermodynamic temperature of the nucleus  under consideration and $I$ is the nuclear thermal moment of inertia. 

The spin distribution $\rho_J$ is determined from the spin projection distribution $\rho_M$ using $\rho_J =\rho_{M=J} - \rho_{M=J+1} \approx -\frac{d \rho_M}{dM} \Big |_{M=J+1/2}$.
Using Eq.~(\ref{M-dist}) leads to the spin distribution in the spin-cutoff model
\be\label{J-dist} 
\frac{\rho_J}{\rho} = \frac{2J+1}{2\sqrt{2\pi} \sigma^3}  e^{-J(J+1)/2\sigma^2}\;.
\ee

The spin-cutoff parameter $\sigma$ given by Eq.~(\ref{sigma-I}) depends on the excitation energy through its dependence on temperature and moment of inertia $I$. At higher excitations, $I$ approaches its rigid-body value~\cite{BM1969}  
\be
I_{\rm rig}/\hbar^2 =\hbar^{-2} \left(\frac{2}{5} M R^2 \right) \simeq 0.0137 A^{5/3} \;\;\;  MeV^{-1} \;,
\ee
but pairing correlations suppresses its value in even-even nuclei at low excitation energies as confirmed in microscopic calculations based on the shell-model Monte Carlo (SMMC)method~\cite{Alhassid2007,Alhassid2005}.  Phenomenological models often assume a moment of inertia at a reduced value at low excitation energies that is interpolated to the rigid-body value at higher energies~\cite{Koning2008}.

Microscopic SMMC calculations showed deviations from the spin-cutoff model in even-even nuclei at low energies in the form of an odd-even staggering effect in spin~\cite{Alhassid2007}, which were confirmed in empirical studies~\cite{vonEgidy2008}.

\subsection{State density versus level density}

The state density is determined by counting the magnetic degeneracy $2J+1$ of each level with spin $J$, i.e.,
\be
\rho(E_x) =\sum_J (2J+1) \rho_J(E_x)  \;.
\ee
This relation is satisfied explicitly by the spin-cutoff formula (\ref{J-dist}) when the sum is replaced by an integral.

Experimentally the spin of measured levels is not always known, in which case energy levels can be counted without including their magnetic degeneracy.  This leads to the level density $\tilde \rho(E_x)$ 
\be \label{eq:level-density}
\tilde\rho(E_x) = \sum_J \rho_J(E_x) \;.
\ee

In the spin-cutoff model, the level density is related to the state density by
\be
 \tilde \rho(E_x)  \approx \frac{1}{\sqrt{2\pi} \sigma}\ \rho(E_x) \;,
\label{Eq:spin-cutoff}
\ee
an expression derived by replacing the sum over $J$ by an integral.

\section{Method to determine state density parameters}\label{method}

The BBF has been used extensively to parametrize NLDs~\cite{Gilbert1965, Dilg1973}.  The BBF for the state densities, Eq.~(\ref{BBF}), contains two parameters, $a$ and $\Delta$ that in the simplest approaches are taken as constants (i.e., independent of the excitation energy).  An additional parameter that only appears in the level density is the spin cut-off parameter $\sigma$, which depends on excitation energy.

In the works of Refs.~\cite{Dilg1973,vonEgidy1988} $a$ and $\Delta$ are adjustable parameters and thus require two experimental data sets to be determined. One data set is the level counting at low energies assuming a complete set of levels is known, and a second data set, when available, is the neutron resonance average level spacing. The neutron resonances provide data at a single excitation energy that equals  the neutron separation energy $S_n$.
The spin-cutoff parameter is calculated from (\ref{sigma-I}) in terms of the nuclear moment of inertia $I$. At higher excitation, the moment of inertia assumes its rigid-body value but at low excitations (i.e., in the counting regime), the moment of inertia $I$ is suppressed by pairing correlations and some modeling of $I$ as a function of excitation energy is required. Ref.~\cite{Dilg1973} carried out fits for two values of the moment of inertia, $I_{\rm rig}$ and  $0.5\, I_{\rm rig}$ for about 200 nuclei with known neutron resonance and level counting data. 

In this work, we use the BBF state density (instead of the BBF level density) which depends only on $a$ and $\Delta$ and does not require knowledge of the spin distribution or of the energy dependence of the spin-cutoff parameter (when the spin distribution is modeled by the spin-cutoff model).

To determine $a$ and $\Delta$ for a given nucleus, we use two experimental data sets: the neutron resonance average level spacing $D_0$ at the neutron separation energy $S_n$ and a complete set of low-lying energy levels including their spin values.

Neutron-capture rates at relatively low energies are dominated by $s$-wave resonances since higher partial waves are suppressed by the Coulomb barrier. The conversion of $D_0$ to state density at $S_n$ requires knowledge of the spin/parity distribution of energy levels. At the neutron separation energy $S_n$, the spin distribution is usually well-described by the spin-cutoff model with rigid-body moment of inertia and the densities of positive and negative parity levels are equal.
Denoting by $I_0$ the spin of the target nucleus, the total spin of the compound nucleus (target nucleus plus a neutron) is $J=1/2$ for $I_0=0$ and  $J=I_0\pm 1/2$ for $I_0 \neq 0$. The experimental state density of the compound nucleus at the neutron separation energy, $\rho_{\rm exp}(S_n)$, can then be inferred from $D_0$ using the relation 
\begin{eqnarray}
\frac{2}{D_0}=
   \left\{ 
   \begin{array}{lrl}
      \rho_{1/2}(S_n) &  & (I_0=0) \\ \rho_{I_0+1/2}(S_n) + \rho_{I_0-1/2}(S_n) &  
      & (I_0 \neq 0) \;,
   \end{array}
   \right.
   \label{Eq:ncapt}
\end{eqnarray}
where  $\rho_J(E_x)$ is the level density of the
compound nucleus with spin $J$ at excitation energy $E_x$.  Assuming a spin-cutoff model (\ref{J-dist}) for $\rho_J$ and a rigid-body moment of inertia in Eq.~(\ref{sigma-I}),  Eq.~(\ref{Eq:ncapt}) determines the total experimental state density $\rho_{\rm exp}(S_n)$ at the neutron separation energy. 
 
The second data set is a complete set of energy levels with known spin values at low excitation energies. Denoting by $N_{\rm exp}(E)$ the cumulative number of experimental states (including the magnetic degeneracy factor of $2J+1$) up to excitation energy $E_x$, the total number of states between a lower excitation energy $E_l$ and a upper excitation energy $E_u$ is
\be
 \int_{E_l}^{E_u} d E' \rho(E') = N_{\rm exp}(E_u)-N_{\rm exp}(E_l) \;,
 \label{Eq:BBFcond}
\ee
where the state density $\rho(E_x)$ on the left hand side is the BBF (\ref{BBF}). At very low excitation energies, the BBF breaks down, especially for positive $\Delta$ where $\rho(E_x)$ has a singularity at $E_x=\Delta$.  The lower energy $E_l$ is chosen such that the BBF is valid down to $E_l$, while the upper excitation energy $E_u$ is chosen so that the spectrum below is complete (including spin values). 

Eq.~(\ref{Eq:BBFcond}) together with  $\rho(S_n)  =  \rho_{\rm exp}(S_n)$, where $\rho(E_x)$ is the BBF (\ref{BBF}) at excitation energy $E_x$, provide a set of two equations for the two unknowns $a$ and $\Delta$. In this work  
we use the discrete level sequence and the average neutron resonance spacings data obtained from the most recent compilation of the Reference Input Parameter Library, RIPL-3~\cite{RIPL}.

 We find that numerical solution to $a$ is rather robust as its value is largely  determined by the value of $\rho_{\rm exp}(S_n)$.  On the other hand, the numerical solution to $\Delta$ is quite sensitive to the chosen energy interval in the level counting regime.  To obtain a suitable energy interval, we vary the lower bound $E_l$ and upper bound $E_u$ of the interval until a satisfactory agreement is reached between the BBF and the experimental state density in the level counting regime (which can
be obtained by binning the states). 
   
\section{Results}

\begin{figure}
	\begin{center}
		\resizebox{0.84\columnwidth}{!}{
			\includegraphics{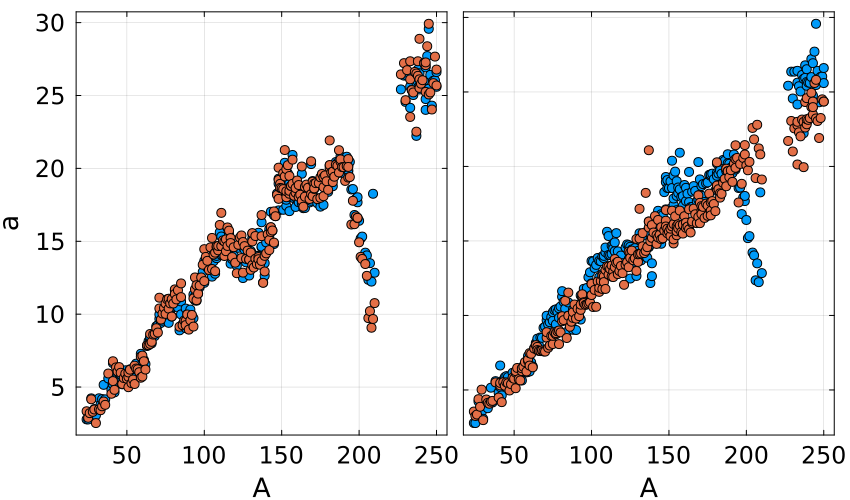} }
		\caption{Level density parameter $a$ vs.~mass number $A$. Our results (blue circles) are compared with those of Ref.~\cite{Koning2008} (red circles). The left panel displays the results of Ref.~\cite{Koning2008} for the parameter $a$  at the neutron resonance energy, i.e., $a(S_n)$, while the right panel shows its asymptotic values, $\tilde{a}$.}
		\label{fig:a_vs_A}       
	\end{center}
\end{figure}

\begin{figure}
	\begin{center}
		\resizebox{0.84\columnwidth}{!}{
			\includegraphics{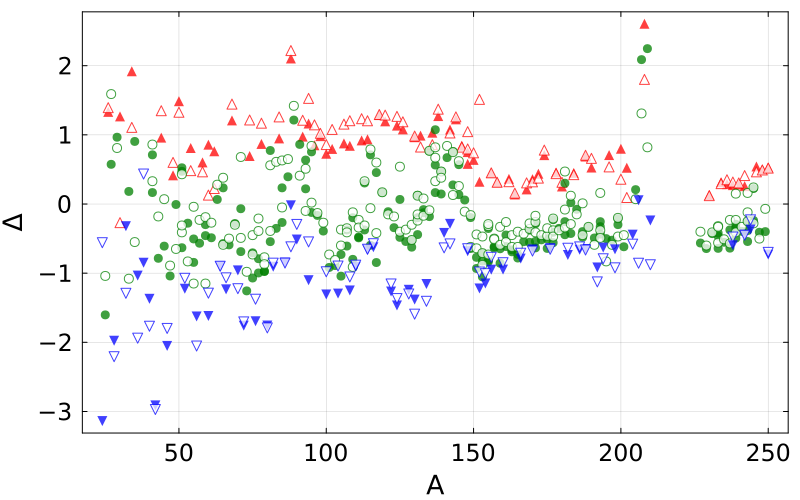} }
		\caption{The backshift parameter $\Delta$  vs.~mass number $A$. Our results (solid symbols) are compared with those of Ref.~\cite{Koning2008} (open  symbols).     Up triangles represent the values for even-even nuclei, circles for odd-mass nuclei, and down triangles for odd-odd nuclei, respectively. }
		\label{fig:Delta_vs_A}       
	\end{center}
\end{figure}

In Table~\ref{Tab:LD}, we present the results of our approach in Sec.~\ref{method} for 294 nuclei, for which neutron resonance data and complete set of  low-lying levels exist. We display the values for the level density parameter $a$ and the back-shift parameter $\Delta$, along with the $l-u$ pairs that we have used to designate the lower and upper indices of nuclear levels used in Eq.~(\ref{Eq:BBFcond}). Fig.~\ref{fig:a_vs_A} and Fig.~\ref{fig:Delta_vs_A} show the distributions of $a$ and $\Delta$ values, respectively, as a function of mass number $A$. Signatures of shell structure are clearly evident in the distribution of $a$, and they become more pronounced for the heavier nuclei. The distribution of $\Delta$ aligns with the systematics of the pairing energies for even-even, odd-even, and odd-odd nuclei while also revealing shell effects. 

We compare our methodology with the local parametrization of the back-shifted Fermi gas model described by Koning~\textit{et al.}~\cite{Koning2008} across 287 nuclei analyzed by both methods. While both approaches share the use of the back-shifted Fermi gas model, the approach of  Ref.~\cite{Koning2008} differs from ours in several significant aspects. Most importantly, it considers in the fits the level density which requires the modeling  of the energy dependence of the spin-cutoff parameter at low excitations while our approach uses the BBF for the state density which is independent of  $\sigma$. In addition, Ref.~\cite{Koning2008} incorporating an energy-dependent level density parameter $a$, and implementing a modification of the BBF proposed by Grossjean and Feldmeier~\cite{Grossjean1985, Demetriou2001} to empirically  address the divergent behavior of the Fermi gas model at very low excitation energies. We note that at higher values of $E_x$, this correction becomes negligible, and the standard form of the BBF is reinstated.

 In Fig.~\ref{fig:a_vs_A} we compare our results for $a$ as a function of $A$ (blue circles) with the results of Ref.~\cite{Koning2008} (red circles). In the left panel we compare our results with those of Ref.~\cite{Koning2008} for $a(S_n)$, the values of $a$ at the neutron separation energy $S_n$. We find close agreement and clear signatures of shell structure. In the right panel we compare our results with the asymptotic values $\tilde a$  of Ref.~\cite{Koning2008}.  We observe discrepancies  around shell closures, as shell effects decay by design in Ref.~\cite{Koning2008}.
  
\begin{figure}
	\begin{center}
		\vspace{-1.0cm}
		\resizebox{1.0\columnwidth}{!}{
			\includegraphics{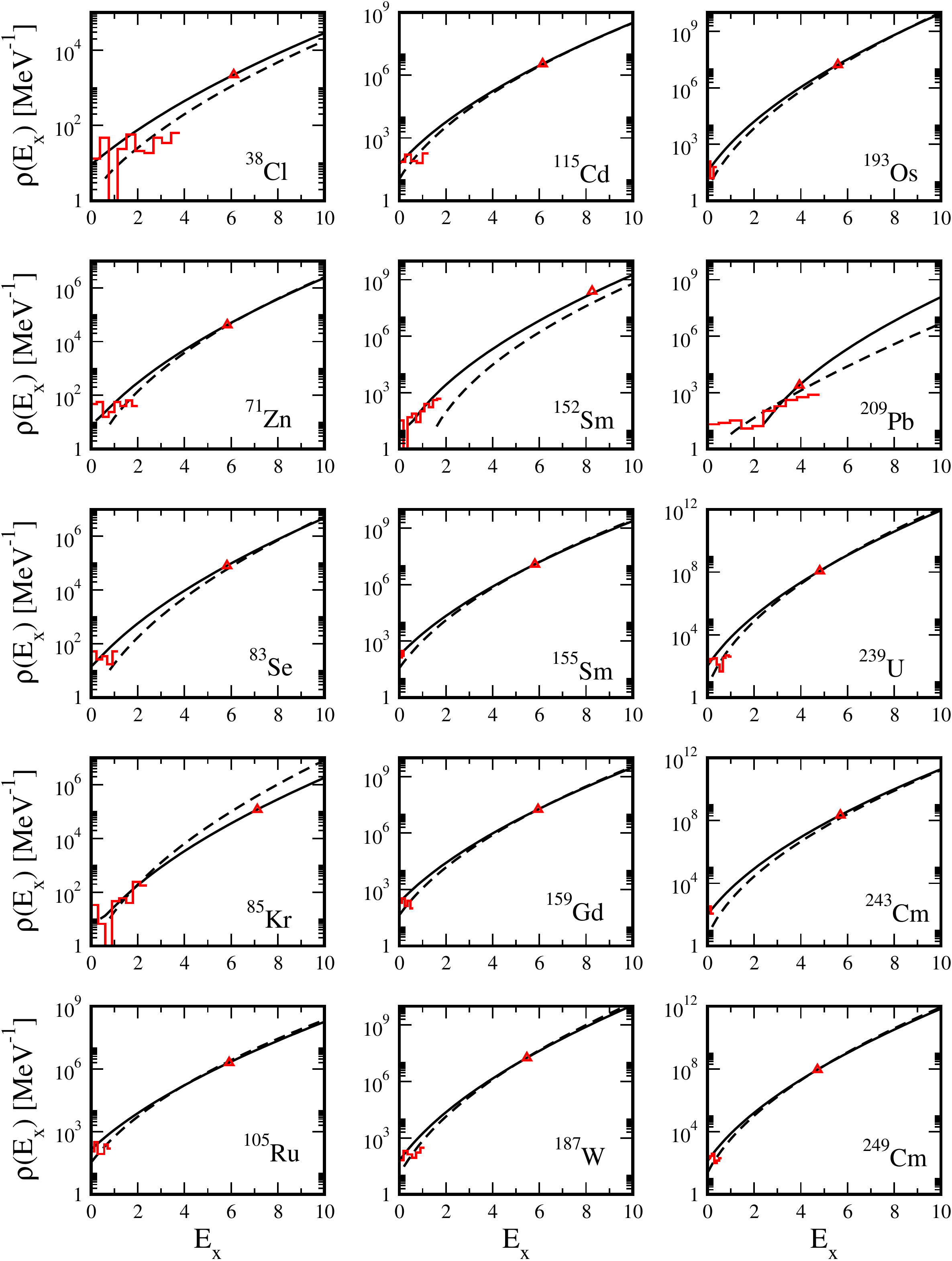} }
		\caption{State densities $\rho(E_x)$ vs.~excitation energy $E_x$, comparing of our results (solid lines) with the locally parametrized back-shifted Fermi gas model of Ref.~\cite{Koning2008} (dashed lines). We show the top 15 nuclei for which the deviations between the two approaches (as measured by max-OMD) are the highest anywhere in the excitation energy interval  $0 \leq E_x \leq 10$ MeV. We also show state counting data (red histograms) and neutron resonance data (red triangles).}
		\label{fig:rho2}       
	\end{center}
\end{figure}
 
In Fig.~\ref{fig:Delta_vs_A}, we compare the two approaches in terms of the distribution of $\Delta$ as a function of mass number $A$. While we observe qualitative agreement between the two approaches regarding mass dependence, odd-even effects, and the signature of shell effects, there are noticeable variations for individual nuclei. These variations are expected, given that the values of $\Delta$ are significantly influenced by low-energy level counting regime, the consideration of which stands as the most noteworthy difference between our methodology and that of Ref.~\cite{Koning2008}.

To assess the deviation between the state density estimates of the two approaches, we calculated the maximum order of magnitude deviation (max-OMD), defined as the maximum value of $\left|\mathrm{log}_{10}\frac{\rho(E_x)}{{\rho}^\prime(E_x)}\right|$ in the excitation energy interval $0 \leq E_x \leq 10$ MeV. Here, $\rho(E_x)$ is the BBF state density in our approach, and ${\rho}^\prime(E_x)$ is the state density according to Ref.~\cite{Koning2008}.

It is important to note that, unlike Ref.~\cite{Koning2008}, we have not employed the Grossjean-Feldmeier modification~\cite{Grossjean1985}. This makes our approach susceptible to divergent behavior at low excitation energies when $E_x$ approaches $\Delta$. To address this, we have limited the comparison of the two approaches to energies where the BBF exhibits monotonic behavior with respect to $E_x$.

We have found the max-OMD values to be below 0.25 for 185 nuclei, between 0.25 and 0.5 for 79 nuclei, and exceeding 0.5 for only 23 nuclei. Fig.~\ref{fig:rho2} illustrates the state densities of the top 15 nuclei possessing the largest max-OMD values anywhere in the interval  $0 \leq E_x \leq 10$. The plot also includes the experimental state densities obtained from state counting data (red histograms) and neutron resonance data (red triangle). Deviations between the two approaches are mostly due to discrepancies at the state counting regime ($^{243}$Cm) or at the neutron resonance ($^{85}$Kr), and in some cases, at both ($^{38}$Cl and $^{152}$Sm). In most of the nuclei shown, we observe that the approach of Ref.~\cite{Koning2008} underestimates the state density at the level counting regime. However, the difference with our state density estimates becomes less significant at higher excitation energies.

\section{Conclusions}

We carried out a local parametrization of the BBF state densities for all 294 nuclei for which neutron resonance data and a complete set of low-lying levels with their spin values exist.  In comparison to other local parametrization of level densities, the novelty of our method is that we determine the two BBF parameters, $a$ and $\Delta$, directly from experimental data using the BBF for state densities.  Other local parameterizations in the literature use the BBF level densities which includes an additional parameter, the spin cutoff parameter $\sigma$, and thus requires a modeling of its dependence on excitation energy. 

Knowledge of the spin distribution is required to convert the measured average neutron resonance spacing $D_0$ to the total state density $\rho_{\rm exp}(S_n)$ at the neutron separation energy. However, at the neutron separation energy it is usually safe to assume a spin-cutoff distribution with a rigid-body moment of inertia.
In contrast, using the BBF for the level density requires the modeling of the dependence of the spin-cutoff parameter $\sigma$ on excitation energy at low excitation energies.   

We tabulated our results for the parameters $a$ and $\Delta$ so that the corresponding BBF state densities can be used to benchmark the experimental results against state densities calculated by various theoretical models.


\section{Acknowledgements}
This work was supported in part by the U.S.~DOE grant No.~DE-SC0019521.

\newpage
\scriptsize
\begin{longtable}{@{\extracolsep{\fill}} rrrrr|lllll}
\caption[State density parameters.]{BBF state density parameters $a$ and $\Delta$, and $l-u$ pairs used in the parametrization (see text). } \label{Tab:LD} \\
\hline
   Z & Nuc  & $l-u$ &  $a$  & $\Delta$ & Z & Nuc  & $l-u$ &  $a$  & $\Delta$ \\
\hline 
11 & ${}^{24}$Na & 4-12 & 2.782 & -3.139  & 30 & ${}^{71}$Zn & 2-11 & 9.981 & -0.068  \\
12 & ${}^{25}$Mg & 6-24 & 2.777 & -1.603  & 31 & ${}^{70}$Ga & 4-10 & 9.248 & -0.960  \\
12 & ${}^{26}$Mg & 6-24 & 3.288 & 1.333  & 31 & ${}^{72}$Ga & 2-10 & 9.403 & -1.759  \\
12 & ${}^{27}$Mg & 4-11 & 4.225 & 0.574  & 32 & ${}^{71}$Ge & 3-7 & 9.678 & -0.688  \\
13 & ${}^{28}$Al & 6-20 & 3.302 & -1.977  & 32 & ${}^{73}$Ge & 1-5 & 9.549 & -1.256  \\
14 & ${}^{29}$Si & 3-16 & 3.489 & 0.966  & 32 & ${}^{74}$Ge & 3-8 & 9.893 & 0.694  \\
14 & ${}^{30}$Si & 4-19 & 3.065 & 1.267  & 32 & ${}^{75}$Ge & 6-10 & 9.640 & -0.706  \\
14 & ${}^{31}$Si & 5-7 & 4.503 & 0.898  & 32 & ${}^{77}$Ge & 3-6 & 9.404 & -0.997  \\
15 & ${}^{32}$P & 3-12 & 4.230 & -0.321  & 33 & ${}^{76}$As & 4-14 & 10.358 & -1.695  \\
16 & ${}^{33}$S & 2-25 & 3.773 & 0.182  & 34 & ${}^{75}$Se & 2-5 & 10.220 & -1.070  \\
16 & ${}^{34}$S & 3-12 & 4.174 & 1.920  & 34 & ${}^{77}$Se & 3-10 & 10.410 & -0.791  \\
16 & ${}^{35}$S & 4-6 & 5.151 & 0.905  & 34 & ${}^{78}$Se & 3-8 & 10.241 & 0.869  \\
17 & ${}^{36}$Cl & 6-9 & 4.157 & -1.035  & 34 & ${}^{79}$Se & 3-9 & 9.841 & -0.974  \\
17 & ${}^{38}$Cl & 4-9 & 5.590 & -0.849  & 34 & ${}^{81}$Se & 8-22 & 11.631 & 0.774  \\
18 & ${}^{41}$Ar & 4-8 & 6.640 & 0.711  & 34 & ${}^{83}$Se & 2-5 & 10.455 & -0.352  \\
19 & ${}^{40}$K & 5-14 & 4.815 & -1.371  & 35 & ${}^{80}$Br & 4-8 & 10.597 & -1.757  \\
19 & ${}^{42}$K & 4-11 & 4.636 & -2.912  & 35 & ${}^{82}$Br & 2-8 & 10.917 & -0.907  \\
20 & ${}^{41}$Ca & 2-7 & 5.193 & 0.166  & 36 & ${}^{79}$Kr & 8-10 & 10.480 & -0.974  \\
20 & ${}^{43}$Ca & 3-13 & 5.664 & -0.785  & 36 & ${}^{81}$Kr & 4-6 & 11.256 & -0.544  \\
20 & ${}^{44}$Ca & 3-8 & 5.857 & 0.960  & 36 & ${}^{84}$Kr & 3-6 & 8.891 & 0.949  \\
20 & ${}^{45}$Ca & 5-10 & 5.927 & -0.609  & 36 & ${}^{85}$Kr & 3-11 & 9.987 & 0.234  \\
21 & ${}^{46}$Sc & 5-11 & 5.861 & -2.052  & 36 & ${}^{87}$Kr & 3-6 & 10.290 & 0.834  \\
22 & ${}^{47}$Ti & 2-6 & 5.186 & -1.041  & 37 & ${}^{86}$Rb & 2-8 & 9.317 & -0.847  \\
22 & ${}^{48}$Ti & 4-6 & 5.560 & 0.415  & 37 & ${}^{88}$Rb & 5-12 & 10.384 & -0.019  \\
22 & ${}^{49}$Ti & 5-13 & 6.150 & -0.007  & 38 & ${}^{85}$Sr & 5-7 & 10.870 & -0.267  \\
22 & ${}^{50}$Ti & 3-11 & 5.712 & 1.487  & 38 & ${}^{87}$Sr & 2-8 & 9.434 & 0.394  \\
22 & ${}^{51}$Ti & 3-16 & 5.942 & 0.473  & 38 & ${}^{88}$Sr & 4-6 & 9.039 & 2.105  \\
23 & ${}^{51}$V & 2-15 & 5.998 & 0.524  & 38 & ${}^{89}$Sr & 3-5 & 9.993 & 1.214  \\
23 & ${}^{52}$V & 5-9 & 6.214 & -1.225  & 39 & ${}^{90}$Y & 2-5 & 9.024 & -0.514  \\
24 & ${}^{51}$Cr & 4-13 & 5.820 & -0.632  & 40 & ${}^{91}$Zr & 4-7 & 10.316 & 0.862  \\
24 & ${}^{53}$Cr & 2-4 & 5.924 & -0.277  & 40 & ${}^{92}$Zr & 3-6 & 9.844 & 0.975  \\
24 & ${}^{54}$Cr & 4-9 & 5.875 & 0.810  & 40 & ${}^{93}$Zr & 3-5 & 11.346 & 0.636  \\
24 & ${}^{55}$Cr & 2-7 & 6.220 & -0.844  & 40 & ${}^{94}$Zr & 3-7 & 11.230 & 1.164  \\
25 & ${}^{56}$Mn & 9-17 & 6.609 & -1.626  & 40 & ${}^{95}$Zr & 3-7 & 11.787 & 0.754  \\
26 & ${}^{55}$Fe & 2-5 & 5.634 & -0.505  & 40 & ${}^{97}$Zr & 2-7 & 11.961 & 0.969  \\
26 & ${}^{57}$Fe & 3-6 & 6.255 & -0.567  & 41 & ${}^{94}$Nb & 3-12 & 10.947 & -1.102  \\
26 & ${}^{58}$Fe & 3-8 & 6.175 & 0.602  & 42 & ${}^{93}$Mo & 5-7 & 9.724 & 0.220  \\
26 & ${}^{59}$Fe & 3-17 & 7.283 & -0.298  & 42 & ${}^{95}$Mo & 2-7 & 10.924 & -0.122  \\
27 & ${}^{60}$Co & 4-10 & 6.929 & -1.619  & 42 & ${}^{96}$Mo & 3-7 & 11.411 & 0.850  \\
28 & ${}^{59}$Ni & 4-8 & 6.222 & -0.457  & 42 & ${}^{97}$Mo & 3-10 & 11.846 & -0.165  \\
28 & ${}^{60}$Ni & 4-10 & 6.241 & 0.858  & 42 & ${}^{98}$Mo & 4-8 & 12.564 & 0.982  \\
28 & ${}^{61}$Ni & 5-12 & 6.852 & -0.488  & 42 & ${}^{99}$Mo & 3-10 & 12.897 & -0.391  \\
28 & ${}^{62}$Ni & 7-9 & 6.577 & 0.763  & 42 & ${}^{101}$Mo & 4-6 & 13.624 & -0.917  \\
28 & ${}^{63}$Ni & 2-8 & 7.817 & 0.066  & 43 & ${}^{100}$Tc & 3-6 & 13.651 & -1.308  \\
28 & ${}^{65}$Ni & 3-5 & 8.269 & 0.234  & 44 & ${}^{100}$Ru & 3-6 & 12.150 & 0.727  \\
29 & ${}^{64}$Cu & 6-12 & 8.147 & -0.902  & 44 & ${}^{101}$Ru & 3-6 & 12.594 & -0.776  \\
29 & ${}^{66}$Cu & 2-7 & 7.996 & -1.234  & 44 & ${}^{102}$Ru & 4-7 & 13.184 & 0.797  \\
30 & ${}^{65}$Zn & 4-13 & 8.442 & -0.572  & 44 & ${}^{103}$Ru & 3-6 & 13.392 & -0.890  \\
30 & ${}^{67}$Zn & 2-6 & 8.523 & -0.591  & 44 & ${}^{105}$Ru & 3-7 & 13.670 & -1.043  \\
30 & ${}^{68}$Zn & 3-9 & 8.499 & 1.209  & 45 & ${}^{104}$Rh & 3-6 & 13.858 & -1.295  \\
30 & ${}^{69}$Zn & 3-6 & 9.169 & -0.282  & 46 & ${}^{105}$Pd & 3-7 & 12.575 & -0.730  \\
\hline \multicolumn{10}{c}{{Continued on next page}} \\ \hline
\end{longtable}

\normalsize
\newpage

\newpage
\scriptsize
\begin{longtable}{@{\extracolsep{\fill}} rrrrr|lllll}
\hline
\multicolumn{10}{c}{{Table \ref{Tab:LD} -- continued from previous page}} \\
\hline
   Z & Nuc  & $l-u$ &  $a$  & $\Delta$ &  Z & Nuc  & $l-u$ &  $a$  & $\Delta$ \\
\hline
46 & ${}^{106}$Pd & 5-8 & 13.634 & 0.877  & 57 & ${}^{139}$La & 3-7 & 12.637 & 0.145  \\
46 & ${}^{107}$Pd & 1-5 & 13.965 & -0.608  & 57 & ${}^{140}$La & 12-18 & 14.923 & -0.418  \\
46 & ${}^{108}$Pd & 4-8 & 13.997 & 0.842  & 58 & ${}^{137}$Ce & 3-9 & 16.304 & 0.165  \\
46 & ${}^{109}$Pd & 1-6 & 14.558 & -0.800  & 58 & ${}^{141}$Ce & 3-9 & 15.439 & 0.671  \\
46 & ${}^{111}$Pd & 5-13 & 15.329 & -0.477  & 58 & ${}^{142}$Ce & 3-11 & 15.828 & 1.077  \\
47 & ${}^{108}$Ag & 2-5 & 13.843 & -1.250  & 58 & ${}^{143}$Ce & 5-14 & 17.025 & 0.276  \\
47 & ${}^{110}$Ag & 2-5 & 15.616 & -0.907  & 59 & ${}^{142}$Pr & 9-16 & 15.244 & -0.287  \\
48 & ${}^{107}$Cd & 2-6 & 12.774 & -0.375  & 60 & ${}^{143}$Nd & 2-25 & 15.951 & 0.759  \\
48 & ${}^{109}$Cd & 6-9 & 14.247 & -0.198  & 60 & ${}^{144}$Nd & 6-12 & 15.719 & 1.225  \\
48 & ${}^{111}$Cd & 2-6 & 14.058 & -0.489  & 60 & ${}^{145}$Nd & 9-12 & 16.792 & 0.135  \\
48 & ${}^{112}$Cd & 4-8 & 14.058 & 0.920  & 60 & ${}^{146}$Nd & 4-9 & 16.724 & 0.843  \\
48 & ${}^{113}$Cd & 2-6 & 14.686 & -0.375  & 60 & ${}^{147}$Nd & 4-20 & 18.276 & 0.058  \\
48 & ${}^{114}$Cd & 3-9 & 14.719 & 0.936  & 60 & ${}^{148}$Nd & 4-11 & 19.293 & 0.581  \\
48 & ${}^{115}$Cd & 3-6 & 14.900 & -0.596  & 60 & ${}^{149}$Nd & 3-8 & 18.552 & -0.644  \\
48 & ${}^{117}$Cd & 2-8 & 13.466 & -0.846  & 60 & ${}^{151}$Nd & 3-7 & 17.159 & -0.932  \\
49 & ${}^{114}$In & 2-9 & 14.487 & -0.661  & 61 & ${}^{148}$Pm & 3-9 & 18.951 & -0.683  \\
49 & ${}^{116}$In & 2-5 & 15.525 & -0.625  & 62 & ${}^{145}$Sm & 5-15 & 14.811 & 0.569  \\
50 & ${}^{113}$Sn & 4-7 & 14.127 & 0.304  & 62 & ${}^{148}$Sm & 3-8 & 17.120 & 0.751  \\
50 & ${}^{115}$Sn & 4-6 & 14.309 & 0.711  & 62 & ${}^{149}$Sm & 6-13 & 19.086 & -0.196  \\
50 & ${}^{117}$Sn & 4-7 & 14.301 & 0.455  & 62 & ${}^{150}$Sm & 4-13 & 18.547 & 0.638  \\
50 & ${}^{118}$Sn & 5-9 & 13.773 & 1.286  & 62 & ${}^{151}$Sm & 5-16 & 18.999 & -0.771  \\
50 & ${}^{119}$Sn & 4-7 & 14.019 & 0.167  & 62 & ${}^{152}$Sm & 3-7 & 18.554 & 0.322  \\
50 & ${}^{120}$Sn & 4-7 & 13.510 & 1.196  & 62 & ${}^{153}$Sm & 3-7 & 17.848 & -1.053  \\
50 & ${}^{121}$Sn & 4-8 & 13.324 & -0.111  & 62 & ${}^{155}$Sm & 2-4 & 17.067 & -0.834  \\
50 & ${}^{123}$Sn & 3-13 & 12.396 & -0.209  & 63 & ${}^{152}$Eu & 3-20 & 20.385 & -1.221  \\
50 & ${}^{125}$Sn & 4-6 & 12.477 & 0.066  & 63 & ${}^{153}$Eu & 2-13 & 17.426 & -0.712  \\
51 & ${}^{122}$Sb & 2-7 & 14.553 & -1.265  & 63 & ${}^{154}$Eu & 3-15 & 19.653 & -1.149  \\
51 & ${}^{124}$Sb & 3-8 & 13.766 & -1.467  & 63 & ${}^{155}$Eu & 5-19 & 18.024 & -0.354  \\
52 & ${}^{123}$Te & 3-6 & 14.767 & -0.342  & 63 & ${}^{156}$Eu & 3-12 & 18.221 & -0.949  \\
52 & ${}^{124}$Te & 5-9 & 14.710 & 1.134  & 64 & ${}^{153}$Gd & 6-15 & 19.399 & -0.825  \\
52 & ${}^{125}$Te & 3-8 & 14.609 & -0.482  & 64 & ${}^{154}$Gd & 5-13 & 16.888 & 0.307  \\
52 & ${}^{126}$Te & 4-6 & 14.430 & 1.079  & 64 & ${}^{155}$Gd & 8-16 & 19.010 & -0.813  \\
52 & ${}^{127}$Te & 3-7 & 13.556 & -0.542  & 64 & ${}^{156}$Gd & 6-9 & 18.588 & 0.449  \\
52 & ${}^{129}$Te & 3-7 & 12.651 & -0.624  & 64 & ${}^{157}$Gd & 3-10 & 18.463 & -0.632  \\
52 & ${}^{131}$Te & 4-15 & 14.149 & 0.126  & 64 & ${}^{158}$Gd & 4-12 & 17.948 & 0.315  \\
53 & ${}^{128}$I & 2-7 & 14.517 & -1.240  & 64 & ${}^{159}$Gd & 3-8 & 17.354 & -0.856  \\
53 & ${}^{130}$I & 12-16 & 14.255 & -1.382  & 64 & ${}^{161}$Gd & 3-6 & 17.677 & -0.448  \\
54 & ${}^{129}$Xe & 4-10 & 13.966 & -0.524  & 65 & ${}^{160}$Tb & 2-7 & 18.756 & -0.948  \\
54 & ${}^{130}$Xe & 4-7 & 14.332 & 0.966  & 66 & ${}^{157}$Dy & 8-15 & 20.923 & -0.616  \\
54 & ${}^{131}$Xe & 3-12 & 15.516 & -0.198  & 66 & ${}^{159}$Dy & 3-9 & 18.008 & -0.647  \\
54 & ${}^{132}$Xe & 3-5 & 13.699 & 0.989  & 66 & ${}^{161}$Dy & 9-19 & 18.564 & -0.615  \\
54 & ${}^{133}$Xe & 3-6 & 13.973 & -0.081  & 66 & ${}^{162}$Dy & 3-15 & 18.172 & 0.364  \\
54 & ${}^{135}$Xe & 6-10 & 15.585 & 0.765  & 66 & ${}^{163}$Dy & 3-8 & 17.605 & -0.663  \\
54 & ${}^{137}$Xe & 3-9 & 14.632 & 0.564  & 66 & ${}^{164}$Dy & 4-10 & 17.269 & 0.138  \\
55 & ${}^{134}$Cs & 1-6 & 13.989 & -1.157  & 66 & ${}^{165}$Dy & 3-9 & 17.284 & -0.728  \\
56 & ${}^{131}$Ba & 3-5 & 15.198 & -0.442  & 67 & ${}^{166}$Ho & 3-12 & 18.683 & -0.789  \\
56 & ${}^{133}$Ba & 6-10 & 15.351 & -0.126  & 68 & ${}^{163}$Er & 4-12 & 19.633 & -0.691  \\
56 & ${}^{134}$Ba & 4-6 & 15.221 & 1.159  & 68 & ${}^{165}$Er & 3-8 & 18.310 & -0.781  \\
56 & ${}^{135}$Ba & 2-6 & 13.999 & -0.182  & 68 & ${}^{167}$Er & 3-8 & 18.357 & -0.545  \\
56 & ${}^{136}$Ba & 4-8 & 13.870 & 1.032  & 68 & ${}^{168}$Er & 5-11 & 17.665 & 0.211  \\
56 & ${}^{137}$Ba & 6-11 & 14.583 & 1.074  & 68 & ${}^{169}$Er & 3-15 & 17.776 & -0.554  \\
56 & ${}^{138}$Ba & 6-10 & 12.176 & 1.272  & 68 & ${}^{171}$Er & 3-11 & 17.935 & -0.540  \\
56 & ${}^{139}$Ba & 4-7 & 13.500 & 0.473  & 69 & ${}^{170}$Tm & 5-13 & 19.132 & -0.656  \\
\hline \multicolumn{10}{c}{{Continued on next page}} \\ \hline
\end{longtable}

\normalsize
\newpage

\newpage
\scriptsize
\begin{longtable}{@{\extracolsep{\fill}} rrrrr|lllll}
\hline
\multicolumn{10}{c}{{Table \ref{Tab:LD} -- continued from previous page}} \\
\hline
   Z & Nuc  & $l-u$ &  $a$  & $\Delta$ &  Z & Nuc  & $l-u$ &  $a$  & $\Delta$ \\
\hline
69 & ${}^{171}$Tm & 3-8 & 18.686 & -0.285  & 80 & ${}^{200}$Hg & 5-13 & 16.407 & 0.799  \\
70 & ${}^{169}$Yb & 7-23 & 20.296 & -0.378  & 80 & ${}^{201}$Hg & 6-8 & 15.188 & -0.618  \\
70 & ${}^{170}$Yb & 6-16 & 18.095 & 0.360  & 80 & ${}^{202}$Hg & 4-9 & 15.323 & 0.525  \\
70 & ${}^{171}$Yb & 3-16 & 18.033 & -0.634  & 81 & ${}^{204}$Tl & 2-7 & 14.220 & -0.443  \\
70 & ${}^{172}$Yb & 6-15 & 18.813 & 0.438  & 81 & ${}^{206}$Tl & 2-5 & 12.359 & 0.054  \\
70 & ${}^{173}$Yb & 3-9 & 17.618 & -0.518  & 82 & ${}^{205}$Pb & 5-8 & 14.032 & 0.208  \\
70 & ${}^{174}$Yb & 8-15 & 18.871 & 0.707  & 82 & ${}^{207}$Pb & 5-19 & 13.497 & 2.088  \\
70 & ${}^{175}$Yb & 4-8 & 17.376 & -0.503  & 82 & ${}^{208}$Pb & 2-15 & 12.234 & 2.607  \\
70 & ${}^{177}$Yb & 4-13 & 17.497 & -0.559  & 82 & ${}^{209}$Pb & 7-26 & 18.260 & 2.247  \\
71 & ${}^{176}$Lu & 7-19 & 19.433 & -0.638  & 83 & ${}^{210}$Bi & 8-17 & 12.842 & -0.238  \\
71 & ${}^{177}$Lu & 6-12 & 19.466 & -0.227  & 88 & ${}^{227}$Ra & 2-8 & 25.419 & -0.564  \\
72 & ${}^{175}$Hf & 2-7 & 18.928 & -0.642  & 90 & ${}^{229}$Th & 5-16 & 26.364 & -0.644  \\
72 & ${}^{177}$Hf & 5-12 & 19.371 & -0.408  & 90 & ${}^{230}$Th & 6-12 & 24.561 & 0.130  \\
72 & ${}^{178}$Hf & 4-8 & 19.515 & 0.457  & 90 & ${}^{231}$Th & 4-15 & 26.363 & -0.428  \\
72 & ${}^{179}$Hf & 2-8 & 19.148 & -0.352  & 90 & ${}^{233}$Th & 4-18 & 26.419 & -0.463  \\
72 & ${}^{180}$Hf & 4-12 & 18.390 & 0.254  & 91 & ${}^{233}$Pa & 3-5 & 24.158 & -0.641  \\
72 & ${}^{181}$Hf & 3-10 & 19.573 & -0.255  & 92 & ${}^{233}$U & 3-12 & 25.623 & -0.420  \\
73 & ${}^{181}$Ta & 20-27 & 20.198 & 0.301  & 92 & ${}^{234}$U & 8-15 & 25.387 & 0.316  \\
73 & ${}^{182}$Ta & 5-10 & 19.524 & -0.737  & 92 & ${}^{235}$U & 4-10 & 25.032 & -0.553  \\
73 & ${}^{183}$Ta & 3-8 & 19.739 & 0.020  & 92 & ${}^{236}$U & 6-10 & 26.141 & 0.288  \\
74 & ${}^{181}$W & 3-8 & 19.371 & -0.444  & 92 & ${}^{237}$U & 4-12 & 25.881 & -0.374  \\
74 & ${}^{183}$W & 3-10 & 18.777 & -0.426  & 92 & ${}^{238}$U & 5-9 & 25.946 & 0.273  \\
74 & ${}^{184}$W & 3-13 & 19.514 & 0.450  & 92 & ${}^{239}$U & 2-5 & 26.527 & -0.340  \\
74 & ${}^{185}$W & 12-18 & 20.437 & -0.077  & 93 & ${}^{237}$Np & 3-9 & 22.231 & -0.638  \\
74 & ${}^{187}$W & 5-8 & 19.844 & -0.399  & 93 & ${}^{238}$Np & 4-10 & 26.788 & -0.596  \\
75 & ${}^{186}$Re & 3-12 & 20.378 & -0.661  & 93 & ${}^{239}$Np & 3-8 & 25.668 & -0.493  \\
75 & ${}^{188}$Re & 3-15 & 20.819 & -0.650  & 94 & ${}^{239}$Pu & 3-9 & 25.413 & -0.292  \\
76 & ${}^{187}$Os & 10-20 & 19.695 & -0.481  & 94 & ${}^{240}$Pu & 4-11 & 25.630 & 0.282  \\
76 & ${}^{188}$Os & 6-9 & 20.211 & 0.687  & 94 & ${}^{241}$Pu & 4-9 & 25.613 & -0.395  \\
76 & ${}^{189}$Os & 11-16 & 20.180 & -0.334  & 94 & ${}^{242}$Pu & 4-6 & 26.000 & 0.259  \\
76 & ${}^{190}$Os & 3-7 & 19.923 & 0.528  & 94 & ${}^{243}$Pu & 5-10 & 26.931 & -0.252  \\
76 & ${}^{191}$Os & 5-22 & 19.556 & -0.491  & 94 & ${}^{245}$Pu & 7-15 & 29.575 & 0.224  \\
76 & ${}^{193}$Os & 2-8 & 19.791 & -0.255  & 95 & ${}^{242}$Am & 4-7 & 27.206 & -0.444  \\
77 & ${}^{192}$Ir & 4-15 & 20.801 & -0.917  & 95 & ${}^{243}$Am & 3-9 & 24.756 & -0.542  \\
77 & ${}^{193}$Ir & 2-9 & 20.144 & -0.327  & 95 & ${}^{244}$Am & 3-7 & 27.708 & -0.369  \\
77 & ${}^{194}$Ir & 3-6 & 20.445 & -0.630  & 96 & ${}^{243}$Cm & 4-8 & 24.014 & -0.384  \\
78 & ${}^{193}$Pt & 2-5 & 19.878 & -0.435  & 96 & ${}^{245}$Cm & 3-8 & 25.465 & -0.237  \\
78 & ${}^{195}$Pt & 2-7 & 18.554 & -0.500  & 96 & ${}^{246}$Cm & 5-10 & 26.415 & 0.543  \\
78 & ${}^{196}$Pt & 4-8 & 18.584 & 0.708  & 96 & ${}^{247}$Cm & 3-7 & 24.305 & -0.408  \\
78 & ${}^{197}$Pt & 3-11 & 16.805 & -0.523  & 96 & ${}^{248}$Cm & 5-10 & 26.015 & 0.500  \\
78 & ${}^{199}$Pt & 3-10 & 18.021 & -0.298  & 96 & ${}^{249}$Cm & 4-10 & 26.039 & -0.399  \\
79 & ${}^{198}$Au & 3-12 & 17.922 & -0.655  & 97 & ${}^{250}$Bk & 3-9 & 26.605 & -0.731  \\
80 & ${}^{199}$Hg & 3-9 & 17.675 & -0.211  & 98 & ${}^{250}$Cf & 5-10 & 25.597 & 0.527 \\
\hline 
\end{longtable}

\normalsize

\newpage

\end{document}